\documentclass[twocolumn,showpacs,pra,aps,superscriptaddress,amssymb]{revtex4}

\usepackage{graphicx}
 
\begin{document}

\title{Delay time and tunneling transient phenomena} 
\author{Gast\'on Garc\'{\i}a-Calder\'on}
\altaffiliation{{\bf Senior Associate ICTP}}
\email{gaston@fisica.unam.mx}
\affiliation{Instituto de F\'{\i}sica,
Universidad Nacional Aut\'onoma de M\'exico,
Apartado Postal {20 364}, 01000 M\'exico, Distrito Federal, M\'exico}
\author{Jorge Villavicencio}
\email{villavics@uabc.mx}
\affiliation{Instituto de F\'{\i}sica,
Universidad Nacional Aut\'onoma de M\'exico,
Apartado Postal {20 364}, 01000 M\'exico, Distrito Federal, M\'exico}
\affiliation{Facultad de Ciencias,
Universidad Aut\'onoma de Baja California,
Apartado Postal 1880, 22800 Ensenada, Baja California, M\'exico}

\date{\today}

\begin{abstract}
Analytic solutions to the time-dependent Schr\"odinger equation for cutoff wave initial conditions are used to investigate the time evolution of the transmitted probability density for tunneling. For a broad range of values of the potential barrier opacity $\alpha$, we find that the probability density exhibits two evolving structures. One refers to the propagation of a {\it  forerunner} related to a {\it time domain resonance} [Phys. Rev. A {\bf 64}, 0121907 (2001)], while the other consists of a semiclassical propagating wavefront. We find a regime where the {\it forerunners} are absent, corresponding to positive {\it time delays}, and show that this regime is characterized by opacities $\alpha < \alpha_c$. The critical opacity $\alpha_c$ is derived  from the analytical expression for the {\it delay time}, that reflects a link between transient effects in tunneling and the {\it delay time}.

\end{abstract}


\maketitle

\section{Introduction}

In recent times there has been relevant technological advances that have made possible to design and construct artificial quantum structures at nanometric scales \cite{qs,corrals}. On the theoretical side the above achievements have stimulated work on the issue of time-dependent tunneling. In particular, one finds a number of works that deal with the solution to the time-dependent Schr\"{o}dinger's equation for cutoff wave initial states \cite{jauho,muga96,gcr97,gcv01}. One interesting feature of these approaches is that at asymptotically long times the time-dependent solution goes into the well known stationary solution. This establishes a bridge between time-dependent and time-independent approaches that may be used to address some subtle questions, such as the controversial problem of the relevant time scales for tunneling\cite{landauer}. 

In a recent work we have used a time-dependent analytic solution to the Schr\"odinger equation for  an arbitrary potential\cite{gcr97}, to explore the tunneling dynamics for a rectangular potential barrier\cite{gcv01}. We found that the probability density exhibits a transient structure that we named {\it time domain resonance}, and obtained that it provides the largest probability of finding the tunneling particle at the potential barrier edge.
Moreover we discussed the relevant time scales associated with the {\it time domain resonance} as a function of the potential parameters and the incidence energy.

The purpose of this work is to extend the above investigation to study  the  time evolution of the probability density along the transmitted region of the potential. We found that for a large variation of potential parameters, the probability density  exhibits two evolving structures. One of them is a {\it  forerunner} that corresponds to the time propagation of the  {\it time domain resonance}, whereas the other structure consists of a propagating wavefront. We find that the {\it  forerunner} vanishes at asymptotically long times and distances from the interaction region, whereas the propagating wavefront tends to the stationary solution of the problem. The propagating wavefront exhibits a time delay with respect to the free case situation. We corroborate that this dynamical {\it delay time} is accurately described by the analytical expression obtained from the phase energy-derivative of the transmission amplitude. The analysis of this time scale as a function of the potential parameters yields positive and negative (time-advance) {\it delay times}. 
We have found that {\it  forerunners} exist whenever the {\it delay time} is negative, thus establishing a deep connection between transient and asymptotic effects in tunneling.
 
The paper is organized as follows: Section II provides the main expressions
of the formalism that are relevant to calculate the probability density
along the transmitted region. In Sec. III we consider the time honored
rectangular barrier potential model. Here we study through several subsections the time evolution of the transmitted probability density, and discuss the {\it delay time}.
Finally, in Sec. IV we present the conclusions.

\section{Formalism}

The relevant expressions to calculate the time evolution of the transmitted
wave with the reflecting cutoff wave initial condition were considered in Ref. \cite{gcv01}. They follow from a general formalism developed by Garc\'{\i}a-Calder\'on \cite {gcr97} for the solution of the time-dependent Schr\"{o}dinger equation for tunneling through an
arbitrary potential $V(x)$ that vanishes outside a region $0\leq x\leq L$. 
Our approach is a generalization of the free case problem considered by Moshinsky \cite{mm} that led to the {\it diffraction in time} phenomenon.   This transient effect has been recently  verified experimentally\cite{dalibard} and has stimulated further studies\cite{zeilinger}.

For the sake of completeness and to fix the notation we recall the relevant
equations here. The cutoff wave initial condition corresponding to a reflecting wave may be written as,
 
\begin{equation}
\psi (x,k;t=0)=\left\{ 
\begin{array}{cc}
e^{ikx}-e^{-ikx}, & x\leq 0 \\[.5cm] 
0, & x>0.
\end{array}
\right.  
\label{2a}
\end{equation}
The time-dependent solution $\psi (x,k;t)$ of Schr\"{o}dinger's equation
for the transmitted region, $x\geq L $ reads, 
\begin{equation}
\psi (x,k;t)=\psi_q(x,k;t) + \psi_r(x,k;t)
\label{3c}
\end{equation}
where $\psi_q$ is given by,
\begin{equation}
\psi_q(x,k;t)= T_kM(y_{k})-T_{-k}M(y_{-k}) 
\label{3d}
\end{equation}
and $\psi_r$ by,
\begin{equation}
\psi_r(x,k;t)=-\sum_n^\infty T_nM(y_{k_n}). 
\label{3e}
\end{equation}
In the above expressions the quantities $T_k$ and $T_{-k}=T_k^*$ refer respectively to transmission amplitudes and $T_{n}=2iku_n(0)u_n(L)\exp(-ik_nL)/(k^2-k_n^2)$ is given in terms of the set of resonant states $\{u_n(x)\}$ and the complex poles $\{k_n=a_n-ib_n\}$ of the problem\cite{gcr97}. The functions $M(y_{s})$ are defined as\cite{gcr97}, 

\begin{equation}
M(y_{s})=\frac 12\mathrm{e}^{(imx^2/2\hbar t)}w(iy_{s}),  
\label{4}
\end{equation}
where the $w(iy_{s})$ is the complex error function \cite{wiz} with the argument 
$y_{s}$ given by 
\begin{equation}
y_{s}\equiv \mathrm{e}^{-i\pi /4}\left( \frac m{2\hbar t}\right)^{1/2}
\left[x-\frac{\hbar s}mt\right] .  
\label{5}
\end{equation}
In Eqs. (\ref{4}) and (\ref{5}), $s$ stands either for $\pm k$ or $k_{\pm n}$, and the index $n$ refers to a given complex pole. Poles are located on the third and fourth quadrants of the complex $k$-plane. The free case solution to the above problem for a cutoff plane wave was considered by Moshinsky \cite{mm}.
The solution for the free case with a reflecting initial condition is given by, 
\begin{equation}
\psi_0(x,k;t)=M(y_{k})-M(y_{-k}).  
\label{5a}
\end{equation}
Note that in the absence of a potential, i.e., $T_k=1$, the
term  $\psi_q$, given by Eq.\ (\ref{3d}), becomes identical to the free case solution $\psi_0$ given above. We shall refer to $\psi_q$, that resembles the free contribution,  as the {\it quasi-monochromatic contribution} and to the sum term given by Eq.\ (\ref{3e}), namely $\psi_r$, as the {\it resonant contribution}.
From the analysis given in Ref. \cite{gcr97} one can see that the exact
solution $\psi=\psi_q+\psi_r$, given by Eq.\ (\ref{3c}), satisfies the initial condition and that at asymptotic long times, $\psi_r \rightarrow 0$ and $\psi_q$ goes into the stationary solution. Hence at very long times $\psi$ becomes, 
\begin{equation}
\psi (x,k;t)=T_k\mathrm{e}^{ikx}\mathrm{e}^{-iEt/\hbar }.  
\label{5b}
\end{equation}

 As pointed out in Ref. \cite{gcv01} a cutoff wave initial state has,
in addition to tunneling components, momentum components that go above the barrier height. One sees from Eq. (\ref{3c}), that the probability density exhibits and interplay between tunneling and over-the-barrier processes. However, as Eq. (\ref {5b}) indicates, at asymptotically long times, the transient effects vanish and one ends up with a stationary tunneling solution.

\section{The Model}

As has been customary in studies involving tunneling times in one dimension
we consider a model that has been used extensively in studies on
time-dependent tunneling, namely, the rectangular barrier potential,
characterized by a height $V_{0}$ in the region $0\leq x\leq L$. To
calculate the time-dependent solution $\psi (x,k;t)$ given 
by Eq.\ (\ref{3c}), in addition to the barrier parameters $V_{0}$, $L$, 
and the corresponding incidence energy $E=\hbar ^{2}k^{2}/2m$, we need to
determine the complex poles $\{k_{n}\}$ and the resonant states $\{u_{n}(x)\}$ of the system. Both the complex poles $\{k_{n}\}$ and the corresponding resonant eigenfunctions $\{u_n(x)\}$, can be calculated using a well established method, as discussed elsewhere. \cite{gcr97, gcv01}.
For the rectangular potential barrier the set of complex energies $E_n=\hbar^2k_n^2/2m =\varepsilon_n-i\Gamma_n/2$ corresponds to the poles of the transmission amplitude of the problem\cite{gcv01}, and hence it may be used to describe the well known top-barrier transmission resonances appearing in that system.

\subsection{Dynamics of the transmitted probability density}

\begin{figure}[!tbp]
\rotatebox{0}{\includegraphics[width=3.3in]{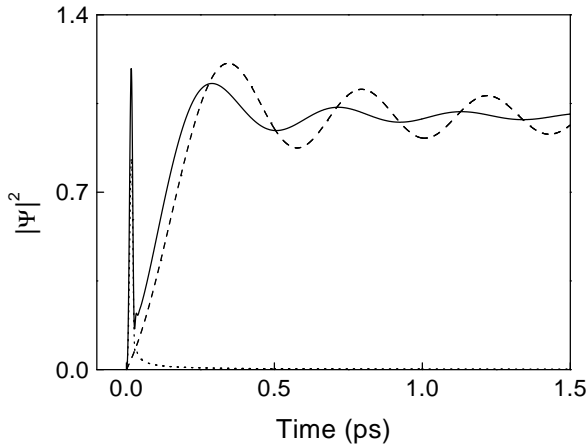}}
\caption{ Time evolution of the normalized probability density $|\psi|^{2}$ (solid line) at the barrier edge $x=L$. The main contribution to the {\it time domain resonance} comes from the resonant term $|\psi_r|^2$ (dotted line), the quasi-monochromatic contribution $|\psi_q|^2$ (dashed line) oscillates with time in a similar fashion as the free solution
$|\psi_0|^2$ \cite{mm} (dashed-dotted line). The system parameters are: $V_0=0.3$ eV, $L=5.0$ nm, and $E=0.01$ eV. See text.}
\label{timedom}
\end{figure}

To exemplify the time evolution of the probability density in the transmitted region we consider a set
of parameters typical of semiconductor artificial quantum structures\cite{qs}: $V_{0}=0.3\,eV$, $L=$ $5.0$ nm, $E=0.01\,$eV and $m=0.067\,m_{e}$, with $m_{e}$ the electron mass. Our choice of parameters is the same as in Ref. \cite{gcv01}, and it guarantees that most momentum components of the initial state tunnel through the potential. 
The different parameters may also be expressed in terms of the opacity
$\alpha$ defined as,
\begin{equation}
\alpha=k_0L,
\label{5ca}
\end{equation}
where $k_0=[2mV_0]^{1/2}/\hbar$, and the ratio $u=V_0/E$. 
In our case $\alpha=3.63$ and $u=30$. 
For example, the regime of  opaque barrier is reached for values of
$\alpha > 5$. In what follows we shall explore the time evolution of $|\psi|^{2}$ at several positions $x_0$ away from the interaction region. From Eq.\ (\ref{3c}) we can write,

\begin{equation}
|\psi|^2=|\psi_q|^2+|\psi_r|^2+I_{rq},
\label{5c}
\end{equation}
where $I_{rq}=2{\rm Re}(\psi_q^*\psi_r)$ stands for the corresponding interference term.
Figure \ref{timedom} displays the time evolution of $|\psi|^{2}$ (solid line) at the right edge of the potential barrier, $x=L$, as considered in Ref. \cite{gcv01}. This is the same example exhibited in Fig. 2 of Ref. \cite{gcv01} using a larger time scale. The sharp peak at very short times is mainly due to the resonant contribution $|\psi_r|^2$ (dotted line). As discussed in Ref. \cite{gcv01}, the maximum value of this {\it time domain resonance}, $t_{p}$, provides a tunneling time scale representing the largest probability to find the particle at the barrier edge. In our example $t_p=5.4$ fs. 
The figure also shows the quasi-monochromatic contribution $|\psi_q|^2$ (dashed-line) that rises and oscillates with time in a manner resembling the free case $|\psi_0|^2$ (dashed-dotted line). 
The interference contribution $I_{rq}$ is not shown, although clearly it is necessary to account for the complete solution.

\begin{figure}[!tbp]
\rotatebox{0}{\includegraphics[width=3.3in]{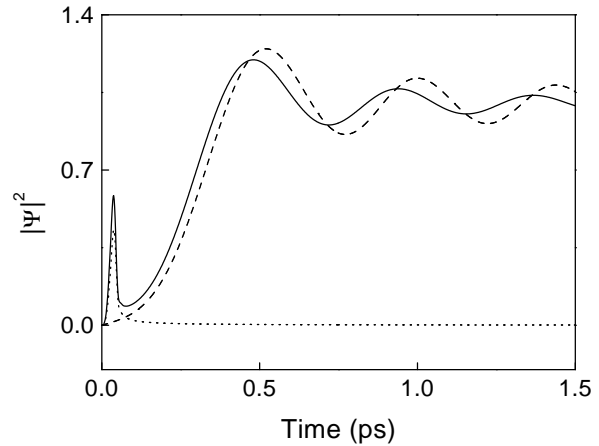}}
\caption{ Time evolution of $|\psi|^{2}$ (solid line) at the fixed position $x_0=50.0$ nm. $|\psi_r|^2$ (dotted line) and $|\psi_q|^2$ (dashed line) account, respectively, by the {\it forerunner} and the quasi-monochromatic contributions. See text.}
\label{tdx50nm}
\end{figure}

Along the transmitted region, $x > L$, the probability density becomes a propagating solution. We can see that the {\it time domain resonance} becomes a propagating structure that we shall refer to as {\it forerunner}. Figure \ref{tdx50nm}  shows the case for $x_0=50.0$ nm. One sees that the amplitude of this transient structure (dotted line) is smaller than the quasi-monochromatic contribution (dashed line). Note also that the solution has separated itself into two well defined structures that propagate with different velocities. The {\it forerunner} propagates with a velocity given approximately by $v_r=\hbar a_1/m$, the velocity associated to the first top barrier resonance,  whereas the quasi-monochromatic contribution does that, approximately by $v_k=\hbar k/m$, the velocity associated to the incident particle.
From a physical point of view we can understand the above situation by noting that our initial state possesses momentum components in k-space above the barrier, which can be transmitted more effectively by the resonance  window corresponding to the first top-barrier resonance. This is the origin of the fast tunneling response, given by the {\it  forerunner}, mainly described by $|\psi_r|^2$. 
In our example, the main contribution to the resonant term, $|\psi_r|^2$, comes from the first top-barrier resonant state. However, depending on the distance $x_0$, one may need to sum up over many terms to account for the complete wave function.
The second type of response is given in a natural way by the maximum of the first peak of the probability density associated with the quasi-monochromatic contribution $|\psi_q|^2$. This  comes mostly from the momentum components centered about the momentum $\hbar k$, which tunnel through the structure.

\begin{figure}[!tbp]
\rotatebox{0}{\includegraphics[width=3.3in]{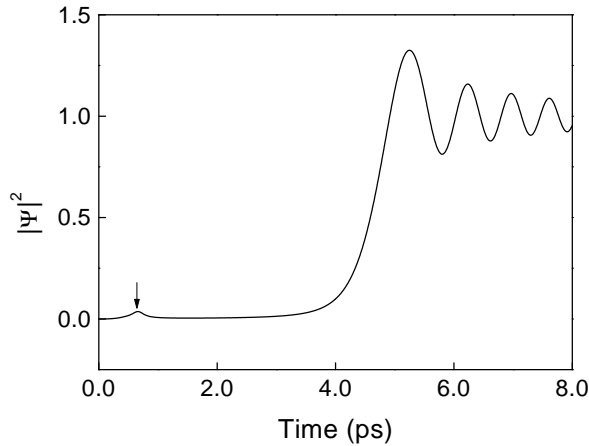}}
\caption{ Time evolution of $|\psi|^{2}$ at the fixed position $x_0=1000.0$ nm.
Notice that the {\it forerunner} has almost disappeared 
(an arrow indicates its position). The parameters are the same 
as in Fig. \ref{timedom}.}
\label{tdx1000nm}
\end{figure}

At still much larger distances, $x_{0}=1000.0$ nm as shown in Fig. \ref{tdx1000nm}, the {\it forerunner} has disappeared almost completely, and the time evolution of the probability density is dominated by $|\psi_q|^2$. 
     
\begin{figure}[!tbp]
\rotatebox{0}{\includegraphics[width=3.3in]{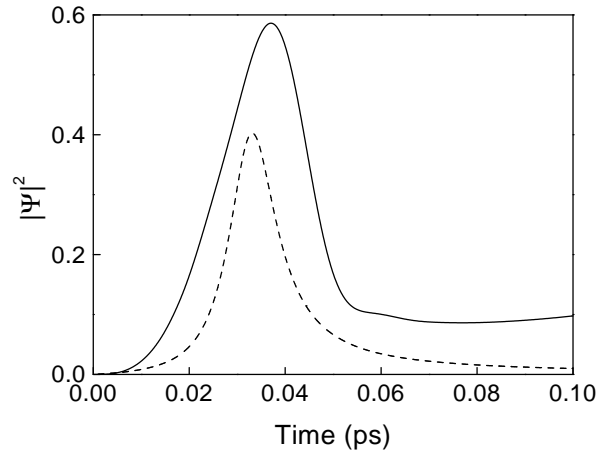}}
\caption{Comparison between the formula for the forerunner, $|\psi_f|^2$
(dashed line), and the exact solution, $|\psi|^2$ (solid line), as a 
function of time for a fixed value of the position $x_0=50.0$ nm.
The parameters are the same as in Fig. \ref{timedom}.}
\label{forerunner}.
\end{figure}

The  behavior of the {\it forerunner} may be understood qualitatively 
by using the asymptotic properties \cite{gcr97} of the $M(y_{k_{\pm n}})$ 
functions  in  Eq. (\ref{3e}). By numerical inspection we find that at a fixed
position $x_0$, the main features of the {\it forerunner} can be  described 
using the one-term ($n=1$) approximation to  Eq. (\ref{3e}), 
namely, $|\psi_f|^2 = |T_1 M(y_{k_1})|^2$. Since in the vicinity of the peak
of the {\it forerunner} the argument $y_{k_1}$ of $M(y_{k_1})$ lies within
$-\pi/2 < arg\,y_{k_1} < \pi/2$, one obtains  
$|\psi_f|^2=(4\pi)^{-1}|T_1/y_{k_1}|^2$. This allows us to write a simple
analytical expression for the time evolution of this transient structure, namely, 
\begin{equation}
\frac{|\psi_f|^2}{|T_1|^2}=\frac{1}{2\pi }\frac{(\hbar t/m)}
{\left[ (x_0-\hbar a_1t/m)^2+(\hbar b_1t/m)^2 \right]}.
\label{tf}
\end{equation}
From the above equation we can see that the peak of the
{\it forerunner} propagates with a velocity $v_r=\hbar a_1/m$, 
as discussed earlier in the text. Figure \ref {forerunner} exhibits a 
plot of $|\psi_f|^2$ (dashed line) as a function of time for the same
parameters used in Fig. \ref{timedom}. We observe  good agreement 
with the exact calculation of $|\psi|^2$ 
(solid line), given by Eq. (\ref{3c}). It is worthwhile to point out 
by inspection of Eq. (\ref{tf}), that as time increases, the maximum 
of the {\it forerunner}, occurring at $\tau=x_0/v_r$, diminishes at 
a rate proportional to $x_0^{-1}$. Hence, for an increasing value of 
$x_0$, the transient structure tends to a vanishing value.

\begin{figure}[!tbp]
\rotatebox{0}{\includegraphics[width=3.3in]{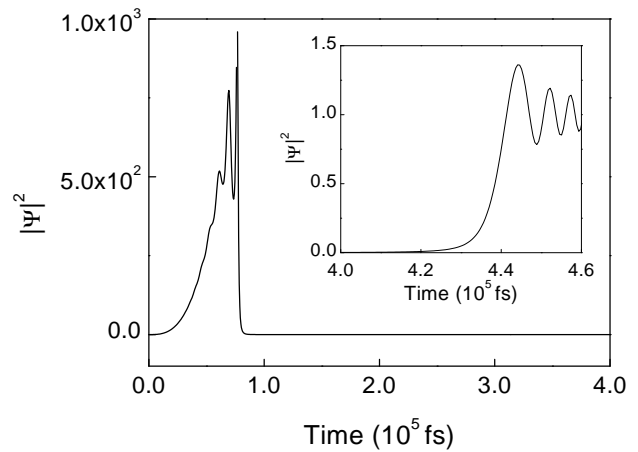}}
\caption{ The main graph shows the time evolution of $|\psi|^2$ for the case of an opaque barrier of width $L=15.0$ nm ($\alpha=10.88$), at a fixed position $x_0=1\times 10^{5}$ nm. Notice that the {\it forerunner}, given essentially by the resonant contribution $|\psi_r|^2$, overwhelms the quasi-monochromatic term described by  $|\psi_q|^2$, as depicted in the inset. See text.}
\label{widebar}
\end{figure}

It is interesting to mention that in the case of opaque barriers, i.e., 
$\alpha > 5.0$, the resonant contribution $|\psi_r|^2$ may be much larger than the quasi-monochromatic contribution $|\psi_q|^2$. This occurs even at quite large distances from the interaction region. Figure \ref{widebar}
exhibits an example of this situation for $L=15.0$ nm and a distance $x_0=1\times 10^{5}$ nm from the potential. Since the solution is normalized to $|T_k|^2$ and this quantity becomes very small for large $L$, one sees that $|\psi_r|^2$ is several orders of magnitude larger than $|\psi_q|^2$,
depicted in the inset to that figure. Clearly, as previously discussed, 
and exemplified in Fig. \ref{tdx1000nm}, at still much larger values of the distance $x_0$, the term $|\psi_q|^2$ shall eventually dominate over $|\psi_r|^2$.
  
\begin{figure}[!tbp]
\rotatebox{0}{\includegraphics[width=3.3in]{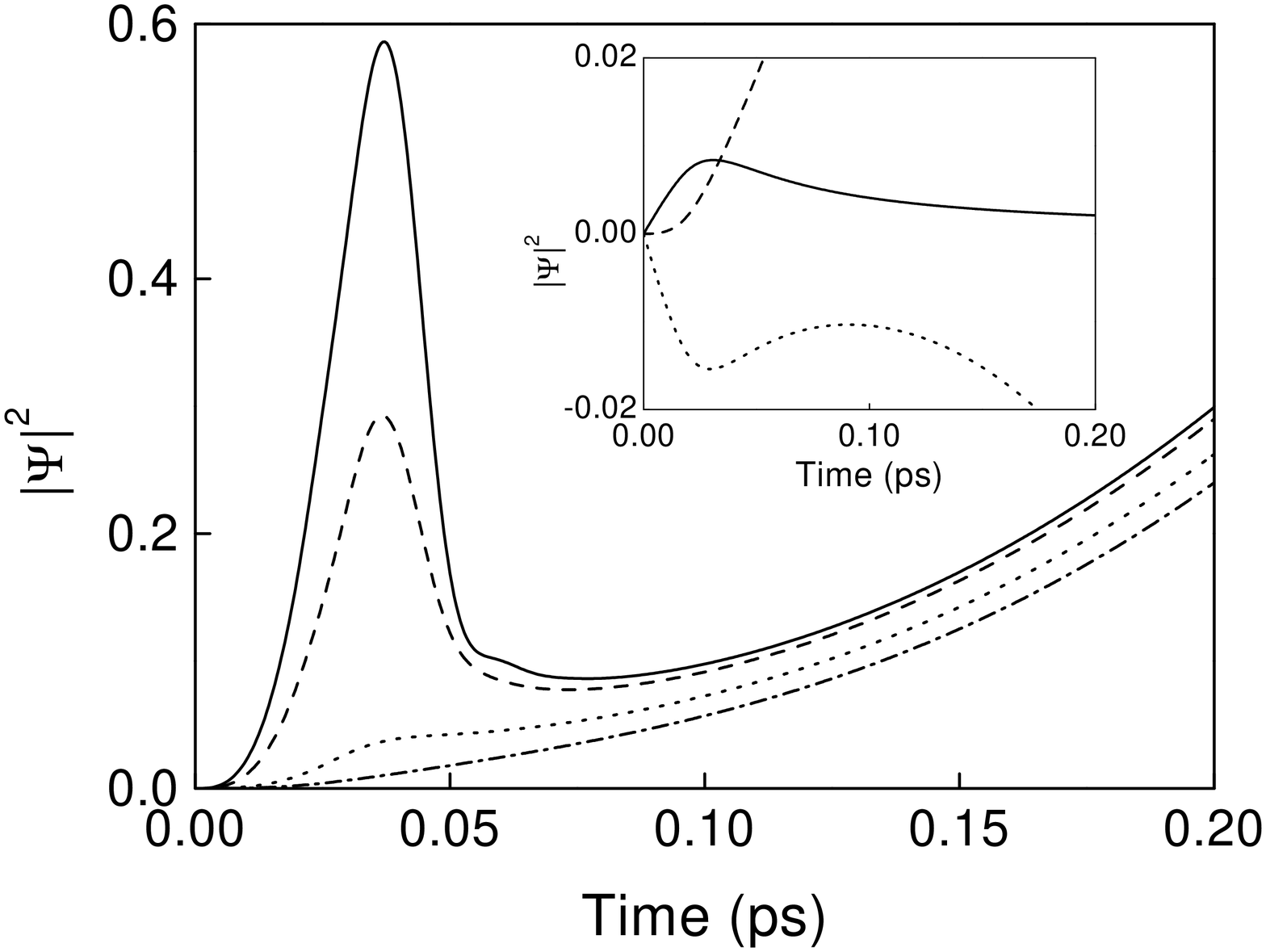}}
\caption{ Time evolution of the \textit{forerunner} for a fixed value of the position $x_0=50.0$ nm, for different values of the barrier width $L$: (a) $5.0$ nm (solid line), (b) $4.5$ nm (dashed line), (c) $3.0$ nm (dotted line), and (d) $2.0$ nm (dashed-dotted line). Notice that the transient structure disappears as the barrier width diminishes. The parameters are the same as in Fig. \ref{timedom}. The inset exhibits a plot of the contributions to  $|\Psi|^2$ (dashed line) of case (d). Notice that the resonant contribution $|\Psi_r|^2$ (solid line) that gives rise to the {\it forerunner}, is almost canceled out entirely by the interference contribution $I_{rq}$ (dotted line).}
\label{vanish}
\end{figure}

Let us now discuss another interesting behavior of the {\it forerunner}. In Fig. \ref{vanish} we plot it as a function of time at a given distance $x_0 > L$,  for different values of the barrier thickness, $L=5.0$ nm (solid line), $L=4.5$ nm (dashed line), $L=3.0$ nm (dotted line), and $L=2.0$ nm (dashed-dotted line), for the same parameters used in Fig. \ref{timedom}. We can see that the intensity of the transient structure diminishes as $L$ decreases. In fact, for the case of a barrier width  $L=2.0$ nm, we observe that the {\it forerunner} disappears. 
However, as shown in the inset to Fig. \ref{vanish}, what happens is that both the resonant contribution, $|\Psi_r|^2$, and the interference term, $I_{rq}$, in Eq.\ (\ref{5c}) for the probability density, are not only overwhelmed by the monochromatic contribution $|\Psi_q|^2$ but also almost cancel each other.

\begin{figure}[!tbp]
\rotatebox{0}{\includegraphics[width=3.3in]{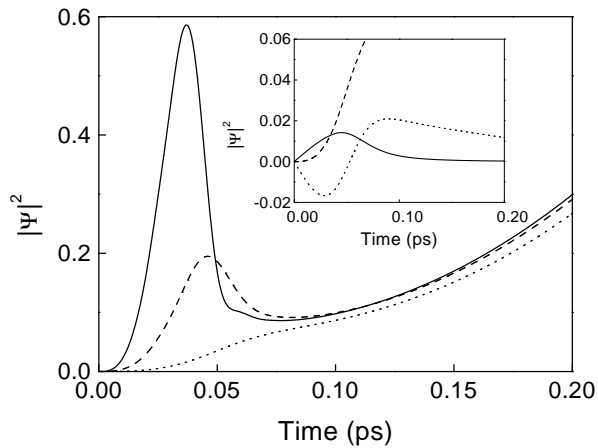}}
\caption{ Time evolution of the \textit{forerunner} for a fixed value of the position $x_0=50.0$ nm, for three different values of the barrier potential $V_0$: (a) $0.3$ eV (solid line), (b) $0.2$ eV (dashed line), and (c) $0.1$ eV (dotted line). Notice that the amplitude of the {\it forerunner} decreases as the barrier height of the potential diminishes.  At the inset we plot the contributions to $|\Psi|^2$ (dashed line) of case (c). Notice that the resonant contribution $|\Psi_r|^2$ (solid line) is almost canceled out by the interference contribution $I_{rq}$ (dotted line).
The parameters are the same as in Fig. \ref{timedom}.}
\label{vanishv0}
\end{figure}

Similarly Fig. \ref{vanishv0} shows that the {\it forerunner} also disappears by diminishing the barrier height $V_0$, and again, as the inset shows, this occurs by the same reason as discussed in the previous case. 
The above results hold also for the {\it time domain resonance}, i.e., at $x=L$. From the above analysis one could argue that the existence of the {\it time domain resonance}, and hence of the {\it forerunners}, depends basically on a particular combination of the parameters $V_{0}$ and $L$. In the next subsection we shall show that this is indeed the case.

\subsection{Delay time and forerunners}

\begin{figure}[!tbp]
\rotatebox{0}{\includegraphics[width=3.3in]{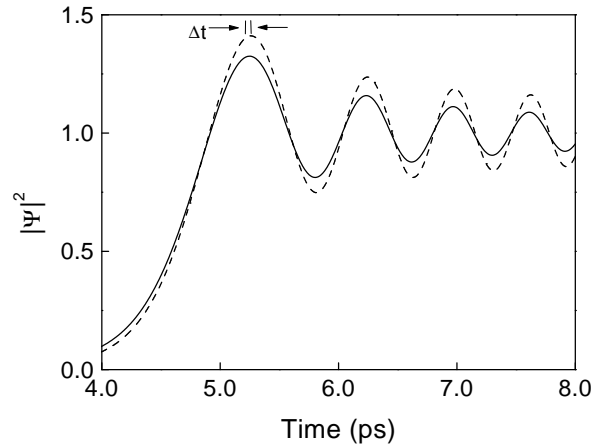}}
\caption{ Time advance of the solution $|\psi|^{2}$ (solid line)
relative to the free propagation case, $|\psi_0|^2$ (dashed line). The parameters are the same as in Fig. 1.}
\label{phase1}
\end{figure}

An interesting result of the analysis of the previous subsection, depicted by Fig. \ref{tdx1000nm}, is that at very large distances from the interaction region the time evolution of the probability density $|\psi|^2$ is essentially given by $|\psi_q|^2$ and exhibits a well defined wavefront. As mentioned above, the  wavefront propagates with approximately the classical velocity $v_k=(\hbar k/m)$, as follows by direct inspection of the argument to the $M-$function, given by Eq.\ (\ref{5}). A comparison of $|\psi|^2$ near the above wavefront  with the corresponding free probability density $|\psi_0|^2$ is exhibited in Fig. \ref{phase1}.
The parameters and the value of the position $x_0$, are the same as in Fig. \ref{tdx1000nm}. Both solutions look very much alike. Note that its corresponding wavefronts are slightly displaced with respect to each other. 
In fact the maximum values of $|\psi|^2$ and $|\psi_0|^2$, for the parameters used in our example, exhibit a time difference that corresponds to a negative {\it delay time} (time-advance).

The above considerations lead us to the notion of {\it delay time} as discussed by Bohm\cite{bohm}. He has argued that the main contribution to the transmitted probability density comes from values in the neighborhood of space, for which the phase of the wave function changes slowly with energy. This yields the well known expression for the {\it delay time}\cite{bohm} as $t_{\phi}=(d\phi/dk)/v_k$ where $\phi$ stands for the phase of the transmission amplitude, i.e., $T_k=|T_k|\exp(i\phi)$ and $v_k$ is the classical velocity as defined above. For the case of the rectangular potential barrier the {\it delay time} reads,

\begin{equation}
t_{\phi}=\frac{m}{\hbar k\kappa}\left[ \frac{k_{0}^{4}\sinh(2\kappa L)-2\kappa Lk^{2}(k^{2}-\kappa^{2})}{4k^{2}\kappa^{2}+k_{0}^{4}\sinh^{2}(\kappa L)}\right] -\frac{mL}{\hbar k}.
\label{6}
\end{equation}

\begin{figure}[!tbp]
\rotatebox{0}{\includegraphics[width=3.3in]{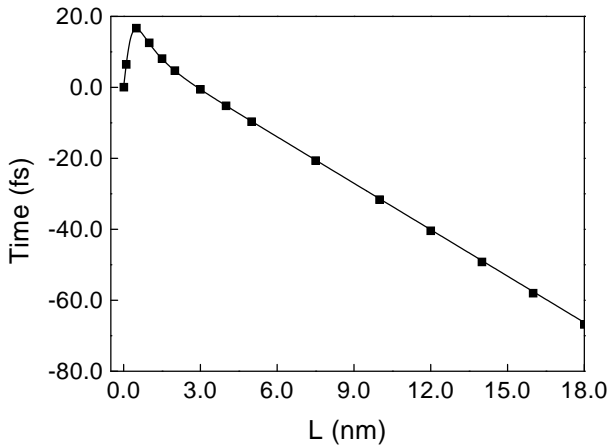}}
\caption{ Plot of $\Delta t$ (full squares) and the {\it delay time} $t_{\phi}$ (solid line)  as a function of the  barrier width $L$, at fixed position $x_0=1\times 10^{5}$ nm, with the same parameters as in Fig. 1. 
Here $k_0=0.07258$ and hence $\alpha$ varies from $7.25 \times 10^{-3}$ to 
$13.065$. See text.}
\label{phasevsL}
\end{figure}

We define $\Delta t$ as the time difference of the maximum values of 
the curves for $|\psi|^{2}$ and $|\psi_0|^{2}$ obtained numerically, 
namely,
\begin{equation}
\Delta t=|\psi^{max}|^{2}-|\psi^{max}_0|^{2}.
\label{7}
\end{equation}
Figure \ref{phasevsL} displays a plot of $\Delta t$ (full squares) and the {\it delay time} $t_{\phi}$ (solid line), as a function of the barrier width $L$ for a large fixed value of the position $x_0$. One sees that $\Delta t$ reproduces exactly the behavior obtained from the analytical expression for $t_{\phi}$. The above agreement of $\Delta t$ with $t_{\phi}$, does not hold when the distance $x_0$ is very close to the interaction region. There, the effect of the transient structure cannot be ignored.  Figure \ref{timedom} exhibits this situation for $x_0=L$. The behavior of $|\psi|^2$ (solid line) is very different from that of the free contribution $|\psi_0|^2$ (dashed-dotted line). As discussed previously the {\it time domain resonance} peak comes from the resonant contribution and hence it is unrelated to the {\it delay time} $t_{\phi}$. The splitting of the solution observed at larger distances and longer times has  yet not occurred. 

Note also in Fig. \ref{phasevsL} that for thin barriers there is a  positive {\it delay time}, and as $L$ increases, there is a transition to a  negative {\it delay time}. In what follows we shall demonstrate that such a transition occurs for a critical value of the opacity $\alpha_c$. In order to show the later, let us rewrite Eq. (\ref{6}) as a function of the parameters $\alpha$ and $u=V_0/E$, namely, 

\begin {equation}
\frac{t_{\phi }}{t_0}=\frac{\left[ 4\gamma^{-1}\sinh \gamma -\cosh \gamma +\gamma^2\alpha^{-2}-3\right]}{\left[ \gamma^2\alpha^{-2}-\gamma^4\alpha^{-4}/4 +\sinh^2(\gamma /2)\right]
\left[ 4-\gamma ^{2}\alpha ^{-2}\right] ^{1/2} },
\label {faseu}
\end {equation}
where we have defined $t_{0}=(mL/\hbar k_{0})$ and $\gamma =2\alpha (1-u^{-1})^{1/2}$.
Thus from Eq. (\ref{faseu}) the condition for the transition from positive to negative {\it delay times}, i.e., $t_{\phi}=0$, is simply given 
by the vanishing of the numerator, namely,

\begin {equation}
4\gamma ^{-1}\sinh \gamma -\cosh \gamma =\left( 3-\frac{\gamma ^{2}}{\alpha ^{2}}\right) 
\label{trascend}
\end{equation} 
For a particular value of the opacity $\alpha$, we can determine from the above equation the value of $u$ at which the transition occurs. 
However, one finds by inspection of Eq.\ (\ref{trascend}), that such a transition is not possible for small values of $\alpha$. That is, there exists a  critical value of the opacity $\alpha= \alpha_c$ such that for 
$\alpha < \alpha_c$, the transition does not occur. This situation corresponds to impose the limit
$u \rightarrow \infty$ in the solution to Eq.\ (\ref{trascend}). This implies that $\gamma \rightarrow 2\alpha_c$, and hence Eq.(\ref{trascend}) becomes,

\begin {equation}
\cosh 2\alpha_c -2 \alpha_c^{-1} \sinh 2\alpha_c =1.
\label{trascend2}
\end{equation} 
The numerical solution to Eq. (\ref{trascend2}) yields the critical opacity $\alpha_c=2.0653$. Note that in addition to the potential parameters $V_0$ and $L$, the opacity depends on the mass $m$ of the incident particle.
It turns out that this value of $\alpha_c$ accounts for systems where the potential barrier is either too shallow or too thin. Therefore, in the regime $\alpha < \alpha_c$, only a positive {\it delay time} is observed.

We have also found that $\alpha_c$ plays an important role in the existence of {\it time domain resonances} and hence of {\it forerunners}. In fact, we find that for systems where $\alpha < \alpha_c$, no {\it time domain resonances} nor {\it forerunners} are observed. This behavior can be seen in Figs. \ref {vanish} and \ref {vanishv0}. For example, in the cases (c) and (d) of  Fig. \ref {vanish} which correspond to barrier widths $L=3.0$ nm and $L=2.0$ nm, the parameter $\alpha$ is, respectively,  2.17 and 1.45. Clearly in case (d) the {it forerunner} has completely disappeared. In Fig. \ref {vanishv0}, the cases (b) and (c) corresponding to the potential heights $V_0=0.2$ eV, and $V_0=0.1$ eV, that refer, respectively, to the values of $\alpha$, 2.96 and 2.095. In this case the disappearance of the {\it forerunner} occurs in the vicinity of the critical opacity $\alpha_c$.
   
\subsection{Comment on the phase-delay time}

Hartman has argued\cite{hartman} that the time it takes to a particle to traverse the classical forbidden region of a potential barrier, can be obtained from an analysis involving the {\it delay time}. He referred to this quantity as the {\it transmission time} $\tau_H$, though nowadays it is often called  {\it phase-delay time}. It corresponds to the difference 
between  the time at which a transmitted particle of momentum $\hbar k$ would leave the rear of the barrier, $x=L$, and the time the same particle would arrive at the front of the barrier, $x=0$. The {\it transmission time} $\tau_H$ can be written as (see Eq. (13) of Ref. \cite{hartman}),
\begin{equation}
\tau_H=t_{\phi}+t_{0},
\label{td}
\end{equation}
where $t_{\phi}$ is given by Eq.\ (\ref{6}) and $t_{0}=(mL/\hbar k)$ represents the free time across a distance equal to the barrier width $L$.
Note that $t_0$ cancels out exactly the second term on the right-hand side 
of Eq.\ (\ref{td}). 
\begin{figure}[!tbp]
\rotatebox{0}{\includegraphics[width=3.3in]{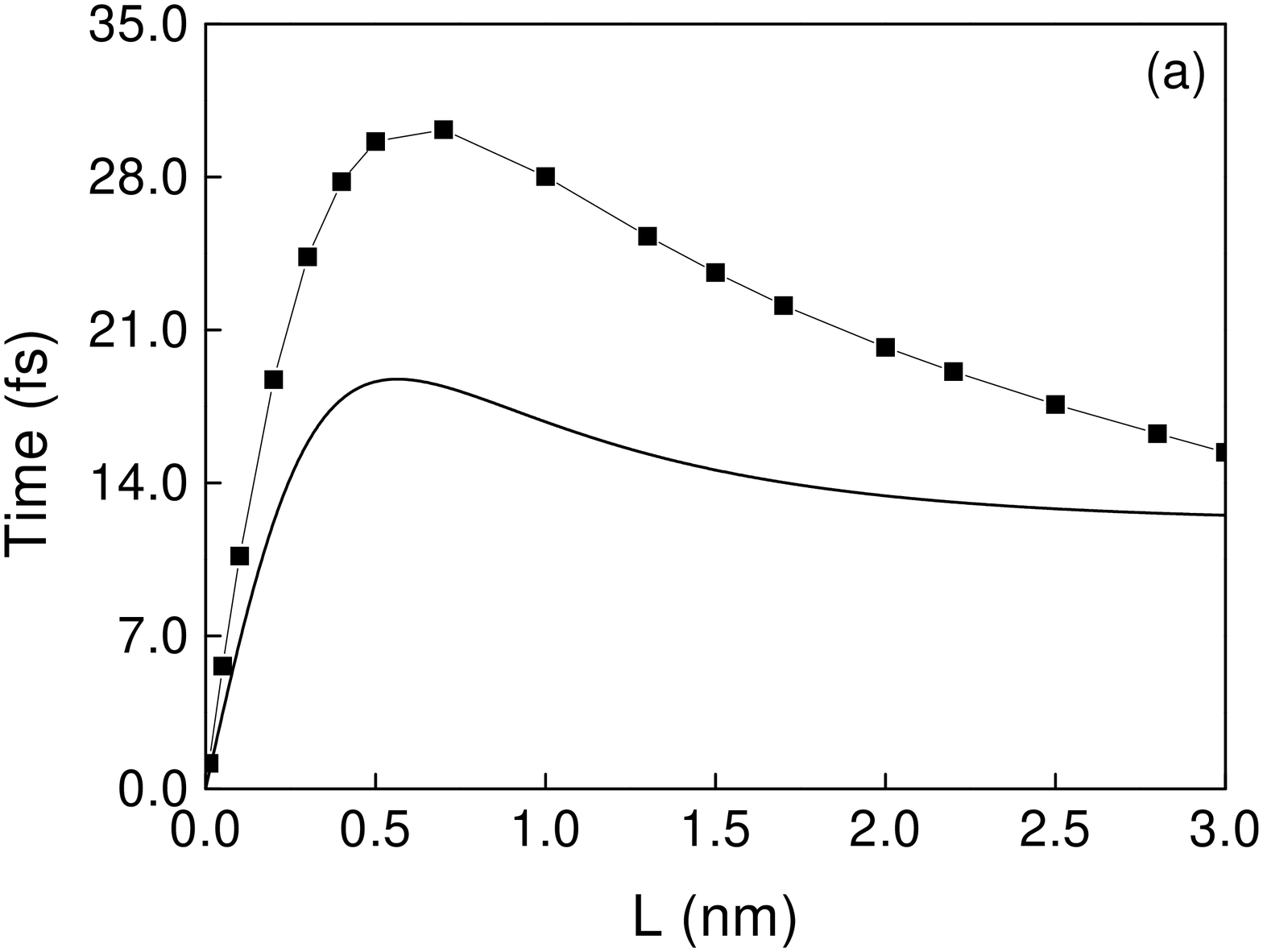}}
\rotatebox{0}{\includegraphics[width=3.3in]{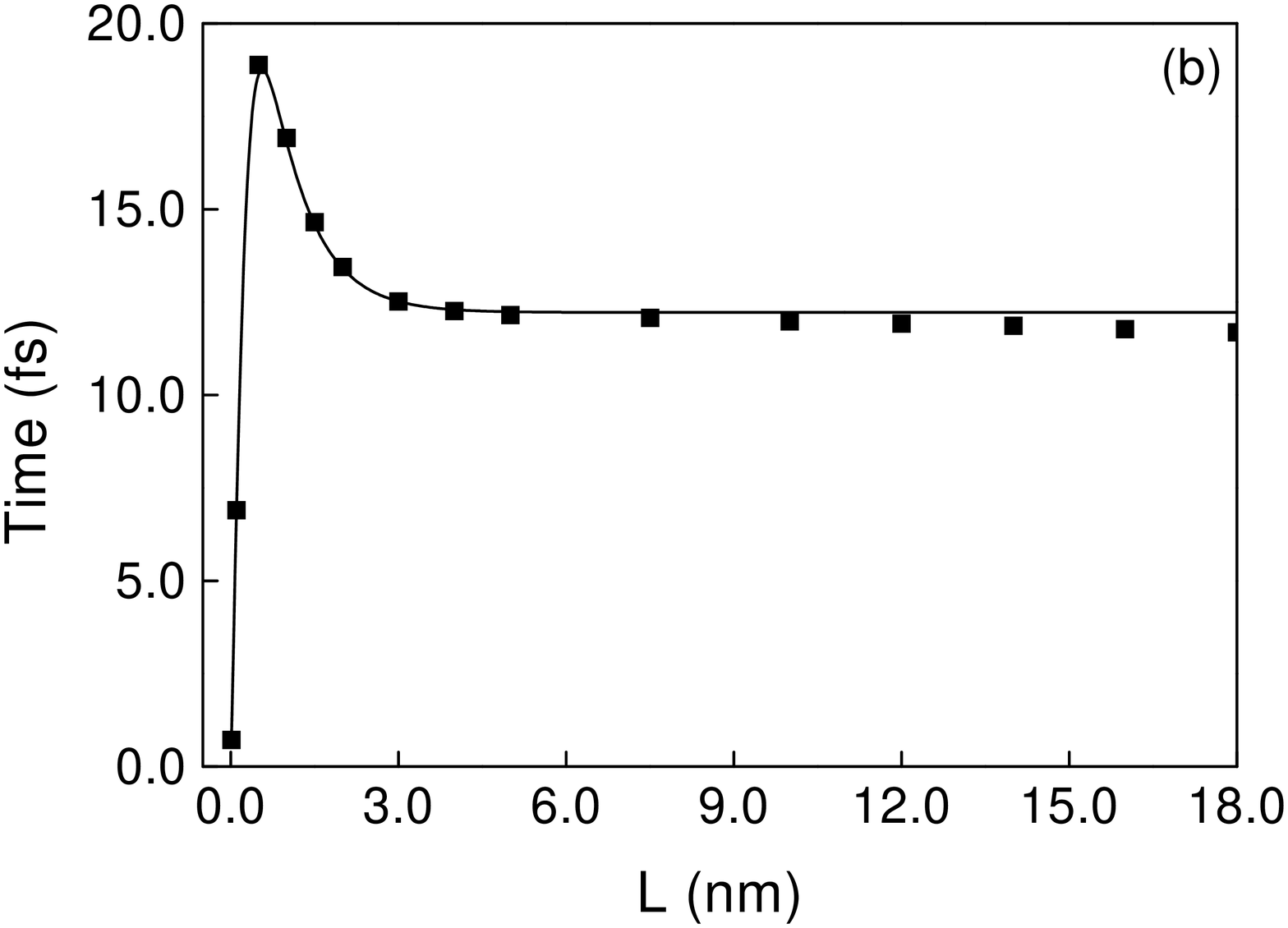}}
\caption{ Comparison of the {\it transmission time} $\tau_H$ (solid line) 
with $\delta_H$ (full squares) as a function of the  barrier width $L$, 
measured at (a) the barrier edge $x=L$ and (b) at a fixed position $x_0=1\times 10^{5}$  nm. The parameters are as in Fig. \ref{timedom}. Hence $k_0=0.7258$, and the opacity varies as mentioned in Fig. \ref{phasevsL}. See text.}
\label{phasevsL1}
\end{figure}
The idea of considering $\tau_H$ as the relevant time scale for tunneling through a classically forbidden region has been criticized by arguing that there is no physical justification for relating in a causative sense the free evolving peak and the transmitted peak through a barrier\cite{buttiker,landauer}. Our analysis of the time evolution of the probability density supports this criticism. Indeed, as discussed in subsection A, along the transmitted region, the probability density may split into two structures evolving  with different velocities. Hence it is not physically justified to choose a feature of one of them to compare it with the free evolving case. In particular, at the barrier edge $x=L$, for $\alpha > \alpha_c$, the behavior of probability density $|\psi(L,t)|^2$ is governed by the {\it time domain resonance}, that yields a completely different time scale\cite{gcv01} than the {\it phase-delay time}. Moreover, even for $\alpha < \alpha_c$, where there is no {\it time domain resonance}, our calculations do not support Hartman's
transmission time $\tau_H$. This is illustrated in Fig. \ref{phasevsL1} (a), where we plot $\tau_H$ as a function of the barrier width $L$ (solid line), and compare it with a plot of $\delta_H= \Delta t+t_0$ (full squares) measured dynamically at the barrier edge $x=L$. 
Although both curves exhibit a similar qualitative behavior, the values of $\tau_H$ and $\delta_H$, are quite different. 
On the other hand, Figure \ref{phasevsL1} (b) exhibits a plot of $\delta_H$
(full squares) as a function of the barrier width $L$,  measured at a distance $x_0$ very far away from the interaction region, i.e., $x_0=1.0 \times 10^{5}$ nm. This figure also shows a  plot of $\tau_H$ (solid line)
and we see that they match quite well for all values of $\alpha$.
The lack of agreement between the plots of $\tau_H$ and $\delta_H$ in 
Fig. \ref{phasevsL1} (a) follows because the {\it time domain resonance}, the quasi-monochromatic  contribution and the interference term are very close together, as exemplified in Fig. \ref{timedom} for $L=5.0$ nm.
On the other hand, at long distances the {\it forerunner}
and the quasi-monochromatic contribution are quite separated, though it may be shown that the interference term accounts for the delay time\cite{gch02}.
 
In the opaque barrier regime, $\alpha \gg 1$, the above times become independent of the barrier width, giving rise to the well known Hartman effect. Indeed at asymptotically large values of $L$, $\tau_H$ goes as $2m/(\hbar k \kappa)$ as follows by inspection of Eq.\ (\ref{td}) \cite{hartman}. As can be seen, it is only at long  distances from the interaction region, when $\delta_H$ coincides with the dynamical time scale $\delta_H$.

The above considerations, therefore, indicate that the {\it transmission time}, i.e., $\tau_H$ given by Eq. (\ref{td}), does not represent the tunneling time of the particle through the classically forbidden region. The {\it phase time delay} $t_{\phi}$, in the spirit of Wigner and Eisenbud\cite{wigner}, represents an asymptotic effect of the potential on the tunneling particle.

\section{Conclusion}
Using an analytical solution to the time-dependent Schr\"{o}dinger
equation for cutoff semi-infinite initial waves, we have investigated the dynamics of the transmitted probability density for tunneling through a rectangular potential barrier. We have found two regimes, characterized by a critical opacity parameter $\alpha_c$, such that for values of $\alpha < \alpha_c$ there are no {\it domain resonances} and consequently no {\it forerunners}, whereas for $\alpha > \alpha_c$, these transient structures may exist depending on the value of $u=V_0/E$. The above result follows from 
an unexpected connection between the existence of these transient structures and the {\it delay time}. This deserves to be further studied. 
An interesting feature of the formalism used in this work is that
it applies to arbitrary potential profiles of finite range. Hence,
the existence of {\it forerunners}, for  a given problem, would depend on the interplay among the different contributions to the probability density given by Eq. (\ref{5c}).
Our results suggest also that the study of transient effects cannot be ignored for a thorough understanding of the tunneling time problem.

\acknowledgments{The authors acknowledge partial financial support of DGAPA-UNAM under grant No. IN101301 and from Conacyt, M\'{e}xico, through Contract No. 431100-5-32082E.}

\end{document}